\documentclass[journal=jacsat,manuscript=article]{achemso}   
\usepackage{amsfonts}
\usepackage{amssymb}
\usepackage{bbold}
\usepackage{graphicx}
\usepackage{amsmath}
\usepackage{tikz}
\usepackage{adjustbox}
\usepackage[normalem]{ulem}
\usetikzlibrary{shapes.geometric, arrows}
\usepackage[english]{babel}
\usepackage{color}
\usepackage[version=3]{mhchem} 
\usepackage{url}
\usepackage[colorlinks,citecolor=black,urlcolor=blue,linkcolor=black]{hyperref} 
\usepackage{threeparttable}
\usepackage{makecell}
\usetikzlibrary{arrows,shapes,positioning,shadows,trees}

\newcommand{\omg}[1]{{\color{red}{}}}

\newcommand{\olp}[1]{{\color{cyan}{}}}

\newcommand{\osvp}[1]{{\color{blue}{}}}

\newcommand{\oyz}[1]{{\color{red}{}}}

\author{Yuzhuo Chen}
\affiliation{Zhejiang Laboratory, Hangzhou 311100, China}
\author{Sebastian V. Pios}
\affiliation{Zhejiang Laboratory, Hangzhou 311100, China}
\author{Maxim F. Gelin}
\affiliation{School of Science, Hangzhou Dianzi University, Hangzhou 310018, China}
\author{Lipeng Chen}
\affiliation{Zhejiang Laboratory, Hangzhou 311100, China}
\email{chenlp@zhejianglab.com}

\title{Accelerating molecular vibrational spectra simulations with a physically informed deep learning model}

\begin{document}


\begin{abstract}
In recent years, machine learning (ML) surrogate models have emerged as an indispensable tool to accelerate simulations of physical and chemical processes. 
However, there is still a lack of ML models that can accurately predict molecular vibrational spectra. 
Here, we present a highly efficient high-dimensional neural network potentials (HD-NNP) architecture to accurately calculate infrared (IR) and Raman spectra based on dipole moments and polarizabilities obtained on-the-fly via ML-molecular dynamics (MD) simulations. The methodology is applied to pyrazine, a prototypical polyatomic chromophore. 
The HD-NNP predicted energies are well within the chemical accuracy (1~kcal/mol), and the errors for HD-NNP predicted forces are only one-half of those obtained from a popular high-performance ML model. 
Compared to the \textit{ab initio} reference, the HD-NNP predicted frequencies of IR and Raman spectra differ only by less than 8.3~cm\textsuperscript{-1}, and the intensities of IR spectra and the depolarizaiton ratios of Raman spectra are well reproduced.  
The HD-NNP architecture developed in this work highlights importance of constructing highly accurate NNPs for predicting molecular vibrational spectra. 
\end{abstract}

\section{Introduction}
Molecular dynamics (MD) simulations provide invaluable insights into many chemical and biological processes. To perform MD simulations, one needs to obtain the forces acting on individual atoms at each integration time step of the simulation. \cite{2002MD} While solving the Schr\"{o}dinger equation directly can yield those forces accurately, the computational cost scales exponentially with the system size. One alternative to alleviate the computational bottleneck is the force-field (FF) method, in which simple empirical functions are used to model relevant interactions and to derive atomic forces analytically.  \cite{1924FF,2011FF,2020HDPES} Because the computational cost of the classical MD increases only linearly with the system size, such FF methods can easily treat the dynamics of systems with many thousands of atoms. However, the classical MD simulations have fundamental limitations on the treatment of electronic polarization effects and chemical reactivity. Their applicability to the accurate description of chemical systems is largely limited by the low accuracy of the underlying FF.  \cite{2015FF}

To circumvent aforementioned deficiencies, the methodology of \textit{ab initio} MD (AIMD) is widely employed. In this method, the motion of the nuclei is described by using quantum mechanical forces computed on-the- fly from the electronic structure method. \cite{1985PRLCPMD,marx2009ab} AIMD provides a versatile tool to accurately simulate a variety of observables, e.g.,  molecular vibrational spectra. \cite{2013aimdspectra}  Infrared (IR) and Raman spectroscopies are  prominent examples of vibrational spectroscopies, \cite{2012irbook,2019ramanbook} the theoretical prediction of which is well established in the community of computational chemistry. \cite{2012calcspectra,2011calcspectra} Contrary to standard static calculation approaches, where anharmonic, conformational, dynamic and temperature effects cannot be accounted for properly,  \cite{2007staticdrawback,2010staticdrawback,2013aimdspectra,2017MLIR}  AIMD can describe anharmonic and dynamic effects and is thus well suitable for the practical simulation of vibrational spectra. \cite{2007staticdrawback,2010staticdrawback} However, while AIMD method can produce high-fidelity results, the high computational cost severely restricts its applications to small systems and short-time dynamics. 

Recently, machine learning (ML) methods, which bridge the gap between the accuracy of \textit{ab initio} electronic structure methods and the efficiency of classical FF methods, offer a promising way to accelerate AIMD simulations.  \cite{2021HDNNP,2021MLFF} Among various ML methods for AIMD simulations, high-dimensional neural network potentials (HD-NNP) based on environment-dependent atomic energies have attracted extensive attentions due to their high efficiency and flexibility in accurate representation of reference data from \textit{ab initio} methods. \cite{2007behler,2014behler,2017behler} For the construction of HD-NNP, a crucial step is the choice of the descriptors. The most commonly used local atomic environment descriptors are formulated using geometric information in terms of distances and angles between pairs and triplets of atoms, such as the atom centered symmetry functions (ACSF), \cite{2011ACSF} the smooth overlap of atomic positions (SOAP), \cite{2013SOAP} the Faber-Christensen-Huang-Lilienfeld (FCHL), \cite{2018FCHL,2020FCHL} the deep potential molecular dynamics (DeePMD), \cite{2018deepmd} and the embedded atomic neural network (EANN). \cite{2019EANN} One disadvantage of those descriptors is that the number of terms included in the model increases with the number of atoms. End-to-end NNPs, on the other hand, take element-dependent embedded atomic types and positions as input and then use message passing neural networks (MPNN) to iteratively update atomic features, which has advantage of reducing complexity as the number of atoms increases. Typical examples of end-to-end NNPs include the crystal graph convolutional neural networks (CGCNN), \cite{2018CGCNN}  PhysNet, \cite{2019PhysNet} SchNetPack,  \cite{schutt2019schnetpack,schutt2023schnetpack} DimeNet, \cite{2020dimenet} and the hierarchically interacting particle neural network (HIP-NN). \cite{2018hipnn} ML-accelerated AIMD simulations are employed to predict accurate IR and Raman spectra with unprecedented computational efficiency. \cite{2017MLIR,2020MLRaman} 

In this work, we introduce an approach that leverages the power of ML technique to accelerate the simulation of IR and Raman spectra by training HD-NNP based on AIMD data. The HD-NNP are constructed by combining computationally efficient modified density-like descriptors \cite{2019EANN} and the MPNN used to iteratively update atomic features. The density-like descriptors contain the interaction terms of triplets of atoms implicitly and thus avoid computationally expensive calculations of triplet terms. 
The trained HD-NNP speed up the MD simulation by a factor of 500, exhibiting extraordinary performance in predicting energy, force, dipole and polarizability of pyrazine molecule well within the \textit{ab initio} accuracy. 
The HD-NNP predicted energies are well within the chemical accuracy, and the errors for HD-NNP predicted forces are only one-half of those obtained from PhysNet, \cite{2019PhysNet} a popular high-performance open-source ML model. 
By analyzing molecular geometries extracted from the MD trajectory generated by HD-NNP, IR and Raman spectra can be obtained using predicted dipole moments and polarizabilities.
These spectra are in good agreement with the \textit{ab initio} reference in terms of frequencies, intensities and depolarization ratios.
It thus demonstrates that constructing high-precision NNP is crucial for obtaining accurate molecular vibrational spectra.

\section{Methodology}

\subsection{Overall architecture of HD-NNP}

\begin{figure}
    \centering
    \includegraphics[width=1\linewidth]{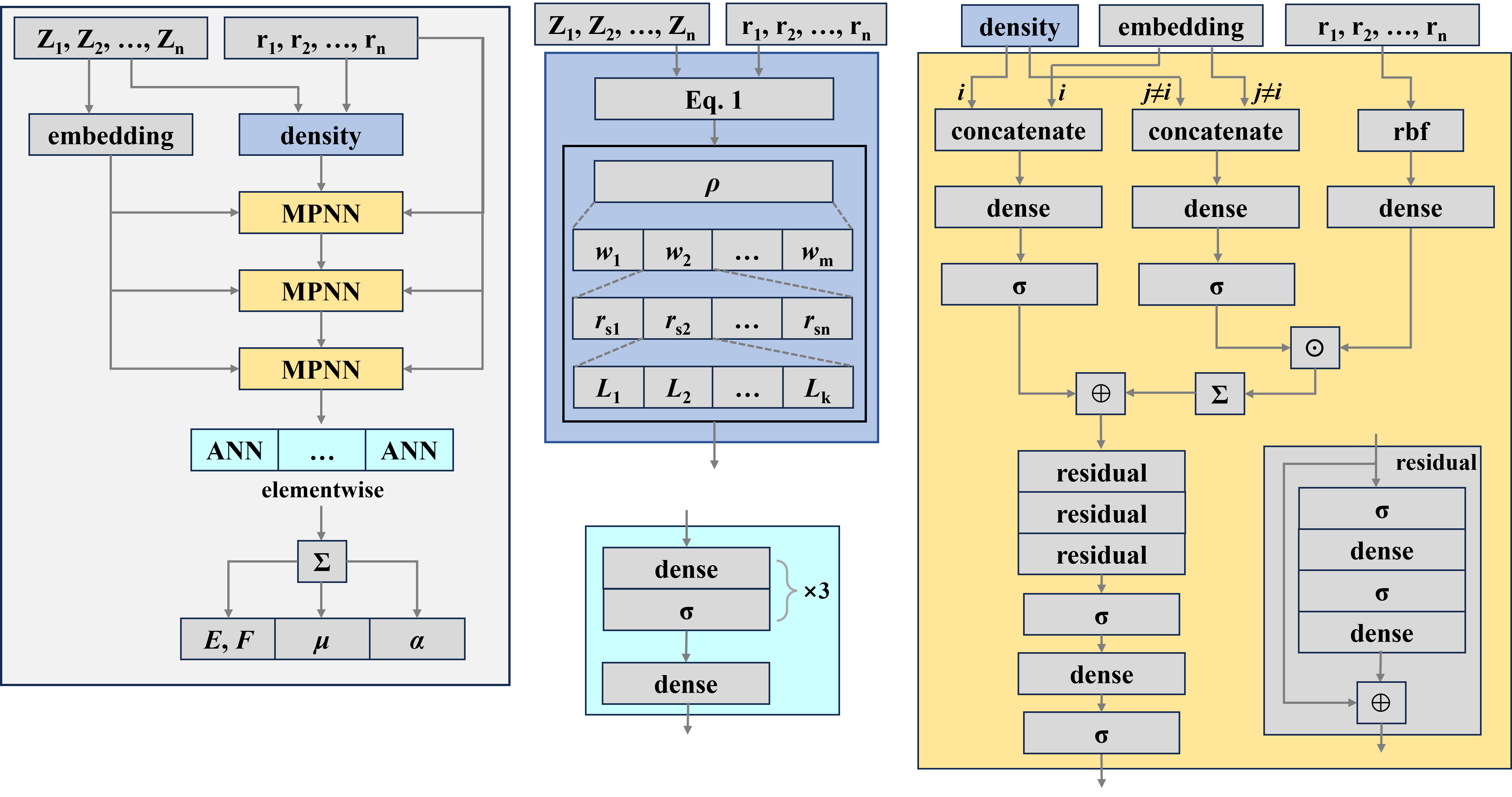}
    \caption{Diagram of the architecture of HD-NNP, in which \textit{Z}\textsubscript{i} and \textbf{r}\textsubscript{i} represent nuclear charges and atomic Cartesian coordinates, respectively. The density is constructed based on Eq.~\ref{eq:density_func} with hyper-parameters $\textit{w}$, $r_\text{s}$ and $L$. Large light gray, cyan, blue and yellow boxes contain the overall architecture of HD-NNP, the structure of ANN, the composition of density-like descriptors and MPNN, respectively. The input nuclear charges \textit{Z}\textsubscript{i} are embedded to feature vectors. The rbf denotes radial basis functions. ``dense", $\sigma$, $\oplus$ and $\odot$ blocks represent the dense layer, the activation function, the direct sum and the Hadamard product, respectively. The output of ANNs are created separately for energies (\textit{E})/forces (\textit{F}), dipole moments ($\mu$) and polarizability ($\alpha$).}
    \label{fig:hd-nnp}
\end{figure}

Figure \ref{fig:hd-nnp} shows the overall architecture as well as key features of HD-NNP developed in this work.
The main architecture is depicted in the large light grey box on the left with detailed illustrations of the density (blue), the atomic neural network (ANN, cyan), and MPNN (yellow) blocks.
The input parameters are atomic charges \textit{Z}\textsubscript{i} and nuclear coordinates \textbf{r}\textsubscript{i}. \textit{Z}\textsubscript{i} are passed into a word embedding layer where they are mapped as randomly initialized and trainable vectors with length 64, as implemented in PyTorch. \cite{pytorch} \textit{Z}\textsubscript{i} and \textbf{r}\textsubscript{i} are jointly used to construct the density embedding. Afterwards, the word and density embedding layers as well as \textbf{r}\textsubscript{i} are passed to MPNN (yellow boxes in Figure \ref{fig:hd-nnp}), which illustrates the key difference of HD-NNP introduced in this work as compared to the previous work.  \cite{2019EANN,2018CGCNN,2019PhysNet}

The embedded electron density of a central atom \textit{i} is described by the summation of Gaussian-type orbitals (GTOs) of neighbouring atoms \textit{j}, weighted by pairwise trainable parameters \textit{c}\textsubscript{ij} and a cutoff function $f_\text{c}(r)$ with a cutoff radius $r_\text{c}$, \cite{2019EANN,1984EAM}

\begin{equation}\label{eq:density_func}
    \rho^i_{\textit{w},r_\text{s},L}(\hat{\textbf{r}}_{\text{ij}})=\sum^{l_\text{x}+l_\text{y}+l_\text{z}=L}_{l_\text{x},l_\text{y},l_\text{z}}{\frac{L!}{{l_\text{x}}!{l_\text{y}}!{l_\text{z}}!}}\left[\sum_j{c_{\text{ij}}\psi^{\textit{w},r_\text{s}}_{{l_\text{x}}{l_\text{y}}{l_\text{z}}}(\hat{\textbf{r}}_{\text{ij}})f_\text{c}(r)}\right]^2,
\end{equation}
\begin{equation}\label{eq:gto}
    \psi^{\textit{w},r_\text{s}}_{{l_\text{x}}{l_\text{y}}{l_\text{z}}}(\hat{\textbf{r}}_{\text{ij}})=x^{l_\text{x}}y^{l_\text{y}}z^{l_\text{z}}\exp{\left[-\textit{w}(r-r_\text{s})^2\right]},
\end{equation}
\begin{equation}\label{eq:cutoff_func}
    f_\text{c}(r)=\left\{
    \begin{aligned}
        (0.5+0.5\cos(\pi{r}/r_\text{c}))^2 \quad r<r_\text{c}\\
        0 \quad r>r_\text{c}\\
    \end{aligned}
    \right..
\end{equation}
Here, $\hat{\textbf{r}}_{\text{ij}} = \hat{\textbf{r}}_{\text{i}}-\hat{\textbf{r}}_{\text{j}}$ denotes the position vector between atoms \textit{i} and \textit{j} with $r=|\hat{\textbf{r}}_{\text{ij}}|$ being the length of $\hat{\textbf{r}}_{\text{ij}}$. \textit{l}\textsubscript{x}, \textit{l}\textsubscript{y} and \textit{l}\textsubscript{z} are orbital angular momenta, whose values range from 0 to \textit{L} and are subjected to the constraint of $l_\text{x}+l_\text{y}+l_\text{z}=L$.
$\textit{w}$, \textit{r}\textsubscript{s}, L are tunable hyperparameters of the density. Different from the previous work, \cite{2019EANN} where element-dependent weights are used for the summation of GTOs, we employ pairwise parameters $c_{ij}$ as weights which allows for more flexible and accurate description of the density. In addition, three-body interaction terms are implicitly included by $L>0$ orbitals as proved by the multinomial theorem. It is also shown that the embedded electron density fulfills the required translational, rotational, and permutational invariances. In the simulation, $\textit{w}$ is fixed at 16.0. \textit{L} are initialized from the sets (0, 1, 2). \textit{r}\textsubscript{s} is initialized from an arithmetic sequence containing 64 elements from 0 to 6~\AA. \textit{r}\textsubscript{c} is fixed at 6.0~\AA. 

The overall architecture of MPNN mimics that of CGCNN\cite{2018CGCNN} and PhysNet\cite{2019PhysNet} (large yellow block in Figure \ref{fig:hd-nnp}). The density and embedded nuclear charges for atom $i$ are concatenated, passed through a dense layer and an activation function. The density and embedded nuclear charges for neighbouring atoms $j$ are also concatenated, passed to another dense layer and activation function, which are followed by the operation of Hadamard product with linear transformed distance expansions using a set of 32 radial basis functions (rbfs). The resultant features are summed and added to the features of atom $i$, which are further passed to a series of residual layers (grey insets in Figure \ref{fig:hd-nnp}). The features from MPNN are then passed to elementwise ANNs, which consist of a series of dense layers and activation functions (large cyan block in Figure \ref{fig:hd-nnp}). Finally, the outputs from each ANN are summed up, yielding separately energies (\textit{E}) and forces (\textit{F}), dipole moments ($\mu$), and polarizability ($\alpha$).

\subsection{Computational details}
Electronic properties of the electronic ground state were calculated using the second-order M\o ller-Plesset perturbation theory (MP2). \cite{mp2_paper} Dunning's augmented correlation-consistent split-valence double -$\zeta$ basis set with polarization functions (aug-cc-pVDZ) on all atoms was employed.  \cite{dunning_basisset} All calculations were accelerated using the resolution-of-identity (RI) approximation.  \cite{riapprox_ref} The MP2/aug-cc-pVDZ method was also used for geometry optimization and subsequent frequency calculations for the vibrational normal modes, which were required to obtain the initial conditions of the AIMD trajectory at 300~K sampled from the Wigner distribution. All electronic structure calculations were performed using Turbomole program package (V7.7). \cite{turbomole}

The dataset was generated by propagating a single AIMD trajectory on the electronic ground state for 25~ps with a time step of 0.5~fs (50,000 data points), including the potential energy (\textit{E}) and force (\textit{F}), the dipole moment ($\mu$), and the polarizability ($\alpha$) at each time-step. The AIMD was performed using ``ZagHop" trajectory surface hopping program package (https://github.com/marin-sapunar/ZagHop) interfaced with Turbomole program package (V7.7). \cite{turbomole} From the AIMD trajectory, 1,000 training points were randomly selected and split into sub-training and validation sets, consisting of 800 and 200 data points, respectively, while the remaining points were held back for testing (49,000 data points). 

We trained three ML models, corresponding to the energy $E$ and the force $F$, the dipole moment $\mu$, and the polarizability $\alpha$, respectively. In the training process, the learning rate was increased linearly from $1.0\cdot 10^{-4}$ to $5.0\cdot 10^{-4}$ within 49 steps based on a warm-up scheme, which was subsequently decreased to $2.0\cdot 10^{-5}$ with a decay rate of $e^{-0.005}$. The batch size and the number of training epochs were fixed to 64 and 100000, respectively.   

We used mean square errors (MSEs) and mean absolute errors (MAEs) to quantify the validity of HD-NNP for the prediction of $E$ and $F$, $\mu$ and $\alpha$. The total loss function for the energy and force, $L_{\mathrm{tot}}$, is defined as 
\begin{equation}\label{eq:total_loss_ef}
L_{\mathrm{tot}}=\frac{L_{\mathrm{energy}}}{2\sigma_1^2}+\frac{L_{\mathrm{force}}}{2\sigma_2^2}+\log(\sigma_1\sigma_2),  
\end{equation}
where $L_{\mathrm{energy}}$ and $L_{\mathrm{force}}$ are loss functions for the energy and force with adjustable uncertainty parameters $\sigma_1$ and $\sigma_2$, respectively. \cite{2018lossfunction} $L_{\mathrm{energy}}$ and $L_{\mathrm{force}}$ can be written as 
\begin{eqnarray}
L_{\text{energy}}&=&\frac{1}{N}\sum^{N}_{i=1}{(E_{\text{pred}}^{(i)}-E_{\text{truth}}^{(i)})^2},  \nonumber \\
L_{\text{force}}&=&\frac{1}{N}\sum^{N}_{i=1}{\frac{\sum_{j=1}^{N_{\text{atom}}^{(i)}}\sum^3_{k=1}(F_{\text{pred}}^{(i)(j)(k)}-F_{\text{truth}}^{(i)(j)(k)})^2}{3N_{\text{atom}}^{(i)}}}.
\end{eqnarray}
Here, $N$ is the total number of data points in a batch, the subscripts ``pred" and ``truth" denote the values predicted by HD-NNP and reference values, respectively. The indices \textit{j} and \textit{k} represent $j$th atom and its three Cartesian coordinates, respectively. The MAEs for the energy and force are defined as 
\begin{eqnarray}
\text{MAE}_{\text{energy}}&=&\frac{1}{N}\sum^{N}_{i=1}{|E_{\text{pred}}^{(i)}-E_{\text{truth}}^{(i)}|}, \nonumber \\
\text{MAE}_{\text{force}}&=&\frac{1}{N}\sum^{N}_{i=1}{\frac{\sum_{j=1}^{N_{\text{atom}}^{(i)}}\sum^3_{k=1}|F_{\text{pred}}^{(i)(j)(k)}-F_{\text{truth}}^{(i)(j)(k)}|}{3N_{\text{atom}}^{(i)}}}.
\end{eqnarray}

The dipole moment predicted by HD-NNP is defined as
\begin{equation}\label{eq:def_dipole}
    \mu_{\text{pred}}= \sum_{j=1}^{N_{\text{atom}}}\mu^{(j)} =\sum_{j=1}^{N_{\text{atom}}}q_j\hat{\textbf{r}}_j^m,
\end{equation}
where $q_j$ is the partial charge of the $j$th atom  modeled by an ANN and $\hat{\textbf{r}}_j^m$ is its distance vector from the molecule’s center of mass. 
Due to it's low value in atomic unit, $\mu$ is represented in unit of $e\cdot$pm, yielding a larger value by a factor of about 189. 
This change reduces the error of trained HD-NNP by about 10\%.
The loss function for $\mu$, $L_{\mathrm{dipole}}$, is calculated as 
\begin{equation}\label{eq:loss_dipole}
L_{\text{dipole}}=\frac{1}{3N}\sum^{N}_{i=1}\sum^3_{k=1}(\mu_{\text{pred}}^{(i)(k)}-\mu_{\text{truth}}^{(i)(k)})^2,
\end{equation}
and the corresponding MAE can be written as 
\begin{equation}\label{eq:mae_dipole} \text{MAE}_{\text{dipole}}=\frac{1}{3N}\sum^{N}_{i=1}\sum^3_{k=1}|\mu_{\text{pred}}^{(i)(k)}-\mu_{\text{truth}}^{(i)(k)}|.
\end{equation}

The molecular polarizability is a $3\times3$ symmetric matrix representing the derivative of the induced dipole with respect to the external electric field. As HD-NNP does not include information about the external electric field, we employed an approach as proposed by Jiang and coworkers \cite{2020TensorEANN} to obtain $\alpha$. To this end, we calculated the derivative of virtual quantities, $p_i$ ($i=1,\cdots,\mathrm{N_{atom}}$), generated by the NN model with respect to atomic coordinates, yielding a  3$\times$\textit{N}\textsubscript{atom} matrix \textbf{D}, 
\begin{equation}\label{eq:def_D}  
\mathbf{D}_j=\sum_{i=1}^{N_{\text{atom}}}\frac{\partial{p_{\text{i}}}}{\partial{\hat{\textbf{r}_j}}}\quad\quad(j=1,\cdots,\mathrm{N_{atom}}).
\end{equation}
The NN-based polarizability tensor $\alpha$ can then be constructed by the summation of three terms, 
\begin{equation}
\alpha=\alpha_1+\alpha_2+\alpha_3,
\end{equation}
where $\alpha_1$ and $\alpha_2$ are defined as 
\begin{eqnarray}
\alpha_1&=&\mathbf{D}(\mathbf{D})^T, \nonumber \\
\alpha_2&=&\hat{\textbf{r}}(\mathbf{D})^T+\mathbf{D}\hat{\textbf{r}}^T,
\end{eqnarray}
and $\alpha_3$ is a scalar matrix introduced to correct the rank-deficiency of $\alpha_1$ and $\alpha_2$ when the molecular geometry is planar. The loss function and MAE for $\alpha$ can be calculated as 
\begin{eqnarray}
L_{\text{polar}}&=&\frac{1}{6N}\sum^{N}_{i=1}\sum^6_{k=1}(\alpha_{\text{pred}}^{(i)(k)}-\alpha_{\text{truth}}^{(i)(k)})^2, \nonumber \\
\mathrm{MAE}_{\text{polar}}&=&\frac{1}{6N}\sum^{N}_{i=1}\sum^6_{k=1}|\alpha_{\text{pred}}^{(i)(k)}-\alpha_{\text{truth}}^{(i)(k)}|.
\end{eqnarray}

The IR spectrum, $I_{\mathrm{IR}}(\omega)$, can be obtained from the Fourier transform of the correlation function for the time derivative of dipole moment as \cite{2013aimdspectra}
\begin{equation}\label{eq:def_IR}
    I_{\text{IR}}(\omega)\propto\int_{-\infty}^{+\infty}\langle\Dot{\mu}(\tau)\Dot{\mu}(t+\tau)\rangle_\tau{e}^{-i\omega{t}}dt.
\end{equation}
Here, $\omega$, $\tau$, \textit{t} are the vibrational frequency, the time lag and the time, respectively. The isotropic and anisotropic parts of Raman spectra, $I_{\mathrm{iso}}(\omega)$ and $I_{\mathrm{aniso}}(\omega)$, can be calculated in a similar way,
\begin{eqnarray}
I_{\text{iso}}(\omega)\propto\frac{(\omega_{\text{in}}-\omega)^4}{\omega}\frac{1}{1-e^{-\frac{\hbar\omega}{k_{\text{B}}T}}}\int_{-\infty}^{+\infty}\langle\Dot{\alpha}_{\text{iso}}(\tau)\Dot{\alpha}_{\text{iso}}(\tau+t)\rangle_\tau{e}^{-i\omega{t}}dt, \nonumber \\
 I_{\text{aniso}}(\omega)\propto\frac{(\omega_{\text{in}}-\omega)^4}{\omega}\frac{1}{1-e^{-\frac{\hbar\omega}{k_{\text{B}}T}}}\int_{-\infty}^{+\infty}\mathbf{Tr}(\langle\Dot{\alpha}_{\text{aniso}}(\tau)\Dot{\alpha}_{\text{aniso}}(\tau+t)\rangle_\tau){e}^{-i\omega{t}}dt,
\end{eqnarray}
where $\omega_{\mathrm{in}}$ is the frequency of the incident light, and $k_B$ and $T$ are Boltzmann constant and temperature, respectively. $\dot{\alpha}_{\mathrm{iso}}$ and $\dot{\alpha}_{\mathrm{aniso}}$ are the time derivative of isotropic and anisotropic components of the polarizability, which are defined as 
\begin{eqnarray}
\Dot{\alpha}_{\text{iso}}&=&\mathbf{Tr}(\Dot{\alpha})/3, \nonumber \\
\Dot{\alpha}_{\text{aniso}}&=&\Dot{\alpha}-\Dot{\alpha}_{\text{iso}}\mathbb{1}.
\end{eqnarray}
Here, $\mathbf{Tr}$ and $\mathbb{1}$ denote the trace operation and the identity matrix, respectively.
$\Dot{\mu}$ and $\Dot{\alpha}$ were obtained using the forward difference quotient between consecutive time steps.
The spectra were plotted using TRAVIS program package. \cite{2012travis,2020travis}

\section{Results and Discussion}

\subsection{The performance of HD-NNP}
Table \ref{tab:benchmark_HDNNP} lists mean absolute errors (MAE) for \textit{E}, \textit{F}, $\mu$ and $\alpha$ predicted by the HD-NNP, the HD-NNP with density component only (denoted as ``density-only"), the HD-NNP with MPNN component only (denoted as ``MPNN-only") and PhysNet. \cite{2019PhysNet} 
The ``density-only" and ``MPNN-only" architectures are depicted in Figure S1 (a) and (b) of the supplementary materials, respectively. An ensemble of 3 models is used for all prediction tasks, which decreases the MAE of energy and force by 30\%, and  dipole and polarizability by 10\%, respectively.

The MAE of \textit{E} and \textit{F} predicted by HD-NNP are 0.0119~kcal/mol and 0.0945~kcal/mol$\cdot$\AA, respectively. The MAE of energy is well within the limit of the chemical accuracy (1~kcal/mol), and the MAE of force is 50\% lower than that predicted by PhysNet, which demonstrates that HD-NNP can well reproduce the AIMD trajectory. As compared to PhysNet, HD-NNP can reduce the MAE of dipole and polarizability by 50\% and 30\%, respectively, illustrating the potential of HD-NNP in accurately reproducing IR and Raman spectra as shown in the next section. In addition, ``density-only" and ``MPNN-only" models are also constructed to gauge the performance of HD-NNP with only density-like descriptors or MPNN. It is found that the ``MPNN-only" model yields MAE for \textit{E}, \textit{F}, $\mu$ and $\alpha$ much lower than those predicted by the ``density-only" model, and adding density-like descriptors to MPNN (the HD-NNP used in this work),  can achieve better predictive performance. 

\begin{centering}
\begin{table}[h]\caption{Mean absolute errors (MAEs) for \textit{E}, \textit{F}, $\mu$ and $\alpha$ as predicted by the HD-NNP, the HD-NNP with density component only (denoted as ``density-only``), the HD-NNP with MPNN component only (denoted as ``MPNN-only``) and PhysNet. An ensemble of 3 models is used for all prediction tasks. The units are given in parenthesis.}
\label{tab:benchmark_HDNNP}
    \centering
    \begin{threeparttable}
    \begin{tabular}{c|c|c|c|c}
        \hline
         & HD-NNP & \makecell[c]{density\\-only} & \makecell[c]{MPNN\\-only} & PhysNet\cite{2019PhysNet} \\ \hline
        \makecell[c]{MAE $E$ (kcal/mol)} & $0.0119$ & $0.184$ & $0.0362$ & $0.0210$ \\
        \makecell[c]{MAE $F$ (kcal/mol$\cdot$\AA)} & $0.0945$ & $1.26$ & $0.314$ & $0.209$ \\
        \makecell[c]{MAE $\mu~(e\cdot{pm})$} & $0.0308$ & $0.143$ & $0.0585$ & $0.0720$ \\
        \makecell[c]{MAE $\alpha_{\text{diag}}~(a.u.)^a$} & $0.0640$ & $0.287$ & $0.123$ & $0.0987$ \\
        \makecell[c]{MAE $\alpha_{\text{offdiag}}~(a.u.)^b$} & $0.0365$ & $0.151$ & $0.0731$ & $0.0562$ \\ \hline
    \end{tabular}
    \label{tab:my_label}
    \begin{tablenotes}
        \footnotesize
        \item[a] Average prediction errors for diagonal elements of the polarizability.
        \item[b] Average prediction errors for off-diagonal elements of the polarizability.
    \end{tablenotes}
    \end{threeparttable}
\end{table}
\end{centering}

\subsection{The IR and Raman Spectra}
We train three models, corresponding to the energy $E$ and the force $F$, the dipole moment $\mu$, and the polarizability $\alpha$, based on the 25~ps AIMD trajectory (see computational details). A new MD trajectory is inferred from the HD-NNP model for $E$ and $F$ (ML-MD), and then the dipole moment and polarizability calculated on-the-fly are used to compute IR and Raman spectra, respectively. It should be noted that the ML-MD simulation only takes about 0.5 hour, 500 times faster than pure AIMD simulation. 

\begin{figure}
    \centering
    \includegraphics[width=1\linewidth]{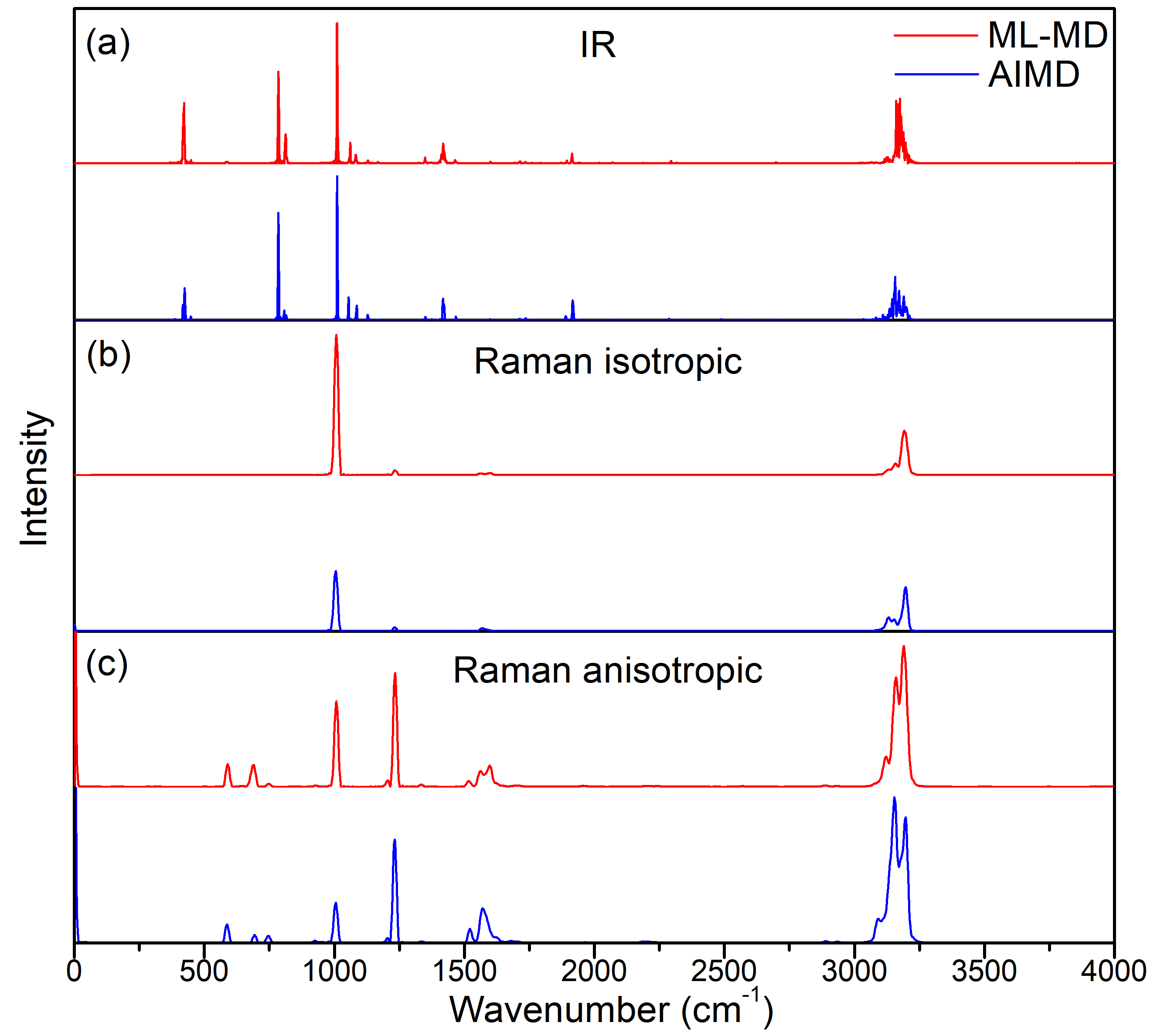}
    \caption{IR (a), isotropic (b) and anisotropic (c) Raman  spectra of pyrazine molecule predicted by the pure AIMD simulation (blue) and the ML-MD simulation (red).}
    \label{fig:IR_Raman_spectra}
\end{figure}

The computed IR, isotropic and anisotropic Raman spectra are displayed in Figure \ref{fig:IR_Raman_spectra} (upper, middle, and lower panel, respectively). Red curves correspond to the results from the ML-MD simulation, while blue curves correspond to pure AIMD results. The frequencies and intensities of IR spectra are listed in Table \ref{tab:IR_peaks}, and the frequencies and depolarization ratios of Raman spectra are listed in Table \ref{tab:Raman_peaks}. The depolarization ratio is defined as the ratio of anisotropic intensity to its isotropic counterpart, which can be related to the symmetry of normal modes of a molecule. A Raman band with a depolarization ratio of less than 0.75 is called a polarized band, while it is called a depolarized band for a depolarization ratio  equal to or greater than 0.75, indicating that a molecule has higher and lower symmetric normal modes, respectively. The IR and Raman spectra can be divided into two regions, normal modes with wavenumbers higher than 3000 $\mathrm{cm^{-1}}$ and lower than 2000 $\mathrm{cm^{-1}}$, corresponding to aromatic C-H stretching modes and other modes, respectively.   

The ML-MD simulation reproduces the peak positions of IR and Raman spectra with maximum deviation of 6.54~cm\textsuperscript{-1} for IR and 8.27~cm\textsuperscript{-1} for Raman, respectively. For IR spectra, the shape of peaks with frequencies higher than 3000 $\mathrm{cm^{-1}}$ predicted by the ML-MD simulation is slightly different from that obtained by pure AIMD simulation. This inconsistency was also observed in other ML models, \cite{2017MLIR} which can be attributed to the difficulty of ML models for the accurate description of C-H stretching motions. It is found that in the frequency regime below 2000 $\mathrm{cm^{-1}}$, the ML-MD simulation overestimate/underestimate the intensity of peaks at 420 $\mathrm{cm^{-1}}$/785 $\mathrm{cm^{-1}}$, respectively. Overall, the agreement between ML-MD and AIMD simulations is quite good.            

\begin{centering}
\begin{table}[h]\caption{List of frequencies (`freq', in unit of  cm\textsuperscript{-1}) and intensities (`int', in unit of \%) of IR spectra obtained from pure AIMD simulation and ML-MD simulation. Peak assignment is based on the results of Billes \cite{1998expRaman} and Chakraborty. \cite{2018peakassign} The sum of intensities of all peaks is set to 100\%.}
\label{tab:IR_peaks}
    \centering
    \begin{tabular}{c|c|c|c|c|c}
        \hline
        mode & AIMD freq & ML-MD freq & $\Delta$ freq & AIMD int & ML-MD int \\ \hline
        $\nu_1(A_u)$ & 424.32 & 420.58 & 3.74 & 8.01 & 10.37 \\
        $\nu_2(B_{3u})$ & 447.13 & 448.06 & -0.93 & 0.87 & 0.89 \\
        $\nu_3(B_{3u})$ & 784.74 & 785.14 & -0.40 & 25.69 & 18.39 \\
        $\nu_4(A_u)$ & 807.02 & 813.02 & -6.00 & 2.57 & 5.48 \\
        $\nu_5(B_{1u})$ & 1011.24 & 1010.71 & 0.53 & 35.47 & 31.19 \\
        $\nu_6(B_{2u})$ & 1054.59 & 1061.13 & -6.54 & 5.20 & 4.35 \\
        $\nu_7(B_{1u})$ & 1128.89 & 1129.02 & -0.13 & 1.08 & 0.45 \\
        $\nu_8(B_{2u})$ & 1350.05 & 1349.78 & 0.27 & 0.69 & 0.92 \\
        $\nu_9(B_{2u})$ & 1417.41 & 1419.15 & -1.74 & 3.97 & 4.52 \\
        $\nu_{10}(B_{1u})$ & 1467.97 & 1464.77 & 3.20 & 0.77 & 0.56 \\
        $\nu_{11}(B_{1u})$ & 3156.30 & 3160.84 & -4.54 & 9.05 & 12.27 \\
        $\nu_{12}(B_{2u})$ & 3171.77 & 3175.11 & -3.34 & 6.58 & 10.55 \\ \hline
    \end{tabular}
    \label{tab:my_label}
\end{table}
\end{centering}

\begin{centering}
\begin{table}[h]\caption{List of frequencies (`freq', in unit of cm\textsuperscript{-1}) and depolarization ratios (`depo') of Raman spectra obtained from pure AIMD simulation and ML-MD simulation. Peak assignment is based on the results of Billes \cite{1998expRaman} and Chakraborty. \cite{2018peakassign}}
\label{tab:Raman_peaks}
    \centering
    \begin{tabular}{c|c|c|c|c|c}
        \hline
        mode & AIMD freq & ML-MD freq & $\Delta$ freq & AIMD depo & ML-MD depo \\ \hline
        $\nu_1(A_g)$ & 587.07 & 589.74 & -2.67 & 0.75 & 0.71 \\
        $\nu_2(B_{3g})$ & 693.15 & 688.61 & 4.54 & 0.75 & 0.75 \\
        $\nu_3(B_{2g})$ & 745.98 & 747.85 & -1.87 & 0.75 & 0.75 \\
        $\nu_4(B_{1g})$ & 925.44 & 929.04 & -3.60 & 0.64 & 0.64 \\
        $\nu_5(B_{2g})$ & 978.41 & 981.48 & -3.07 & 0.043 & 0.045 \\
        $\nu_6(A_g)$ & 1005.36 & 1007.76 & -2.40 & 0.042 & 0.044 \\
        $\nu_7(A_g)$ & 1232.05 & 1233.52 & -1.47 & 0.52 & 0.54 \\
        $\nu_8(B_{3g})$ & 1335.99 & 1334.39 & 1.60 & 0.74 & 0.74 \\
        $\nu_9(B_{3g})$ & 1521.59 & 1517.45 & 4.14 & 0.73 & 0.74 \\
        $\nu_{10}(A_g)$ & 1570.42 & 1562.15 & 8.27 & 0.38 & 0.38 \\
        $\nu_{11}(B_{3g})$ & 3154.32 & 3159.65 & -5.33 & 0.39 & 0.37 \\
        $\nu_{12}(A_g)$ & 3196.61 & 3189.54 & 7.07 & 0.15 & 0.17 \\ \hline
    \end{tabular}
    \label{tab:my_label}
\end{table}
\end{centering}

For Raman spectra, the ML-MD simulation can accurately predict the depolarized ($\nu_1$, $\nu_2$, $\nu_3$, $\nu_8$, $\nu_9$) and polarized bands. In the frequency region below 2000 $\mathrm{cm^{-1}}$, ML predicted depolarization ratios match the pure AIMD ones quite well, with the largest relative error at around 5\%. The relative errors of ML predicted depolarization ratios in the frequency region above 3000 $\mathrm{cm^{-1}}$ slightly increase due to the inappropriate description of aromatic C-H stretching motions. Both isotropic and anisotropic Raman spectra exhibit the largest difference of peak intensities for $\nu_6$ and $\nu_{11}$ modes, with the former overestimated and latter underestimated, respectively. However, the corresponding depolarization ratios are accurately predicted. Overall, ML models developed in this work are capable of accurately reproducing both isotropic and anisotropic Raman spectra as well as depolarization ratios of  pyrazine molecule.  

We also use HD-NNP to predict the dipole moment and polarizability for the geometries of pyrazine molecule from the AIMD trajectory (in the following referred to as ``AI-AIMD"). The computed IR and Raman spectra match almost exactly with those from the pure AIMD simulation (see Figure~S2), demonstrating the importance of the accurate description of forces by ML models when simulating vibrational spectra. Subsequent work could be focused on constructing better ML models to obtain highly accurate ML-predicted trajectory.  

\section{Conclusions}
In this work, we develop a novel architecture of neural-network potentials, the HD-NNP, to predict energies, gradients, dipole moments and polarizability based on the AIMD trajectory. We test the performance of HD-NNP by calculating IR and Raman spectra of pyrazine molecule in the gas-phase, yielding results in good agreement with pure AIMD simulations. It is shown that the prediction of forces with high accuracy is crucial to reproduce vibrational spectra, which is very challenging for the ML-accelerated long-time AIMD simulation. It is thus imperative to design highly accurate ML models to further reduce predictive errors of forces or to predict long-time dynamics based on the short-time trajectory. Work in these directions is in progress.
\begin{acknowledgement}
Y.Z.C, S.V.P. and L.P.C. acknowledge support from the Key Research Project of Zhejiang Lab (No. 2021PE0AC02). M. F. G. acknowledges support from the National Natural Science Foundation of China (No.~22373028).
\end{acknowledgement}

\bibliography{main}

\end{document}